\documentclass{kluwer}    

\usepackage[]{psfig}

\newdisplay{guess}{Conjecture}

\begin{document}  

\begin{article}
\begin{opening}         
\title{Hydrogen-poor planetary nebulae} 
\author{Albert A. \surname{Zijlstra}}  
\runningauthor{Albert Zijlstra}
\runningtitle{Hydrogen-poor PNe}
\institute{UMIST, department of Physics, P.O. Box 88, Manchester M60 1QD, UK}

\begin{abstract}
Five planetary nebulae are known to show hydrogen-poor material near
the central star. In the case of A58, this gas was ejected following a
late thermal pulse similar to Sakurai. In this paper I will review
these five objects. One of them, IRAS 18333$-$2357, may not be a true
PN. I will show that there is a strong case for a relation to the [WC]
stars and their relatives, the weak emission-line stars.  The surface
abundances of the [WC] stars are explained via diffuse overshoot into
the helium layer. The hydrogen-poor PNe do not support this: their
abundances indicate a change of abundance with depth in the helium
layer. A short-lived phase of very high mass loss, the r-AGB, is
indicated.  Sakurai may be at the start of such a phase, and may
evolve to very low stellar temperatures.
\end{abstract}

\end{opening}           

\section{Introduction}  

Planetary nebulae (PNe) are the evidence for a phase of stellar
evolution where the stellar wind reaches $10^{-6}$--$10^{-4}\,\rm
M_\odot yr^{-1}$ at a velocity of 10--$20\, \rm km\, s^{-1}$. Such
winds occur on the Asymptotic Giant Branch (AGB). Abundances in AGB
winds can be affected by carbon dredge-up and s-process enrichment, but
otherwise reflect normal ISM values.

Five PNe are known to have hydrogen-poor inner regions.  This material
cannot have been ejected on the AGB but must trace post-AGB or
pre-White Dwarf evolution. The central stars have therefore
experienced a second phase of high mass-loss rates at low outflow
velocity. Normal post-AGB evolution predicts that the surface of the
star retains a H-rich layer with mass of $\sim 0.01\,\rm M_\odot$
which shields the underlying H-poor layers.  The favoured model for
removing these layers involves a late thermal pulse (Herwig, these
proceedings), as now shown by Sakurai's object. Therefore, it seems
likely that Sakurai's object will become the 6th member of the
class. 

There are in fact H-poor post-AGB stars known, in particular the R Cor
Bor stars, the extreme helium stars, and the [WC] central stars of
PNe. The central stars of the H-poor PNe, and by implication Sakurai
as well, may be evolutionary related to one (or more) of these
classes.

In this review paper I will first discuss the 5 PNe with hydrogen-poor
inner shells.  I will discuss the morphologies and
kinematics. Abundances and dust will be discussed. The possible
relation to the [WC] stars will be discussed. Finally the relation to
Sakurai's object will be discussed.  An earlier review can be found in
Harrington (1996).

\section{Numbers and birth rates}

H-poor material in PNe is identified through long-slit spectroscopy
and/or narrow-band imaging, searching for regions which have very
large ratios of [OIII]/H$\alpha$ or [NII]/H$\alpha$. The H-poor
regions are expected to be embedded in the outer H-rich regions.
Detection requires sufficient spatial resolution to separate the
regions. High contrast is also a plus, since the H-rich and H-poor
regions are seen superimposed. This favours old, low
surface-brightness nebulae where the recent ejecta stand out clearly
against the faint outer nebulosity.

It is therefore no surprise that three of the five H-poor PNe are
Abell-type nebulae: old, extended and faint. Similar regions in young,
bright PNe could easily have gone unnoticed. In total there are 72
Abell-type PNe known, of which three (4\%) contain H-poor material.
If we assume that the sample of H-poor Abell-type nebulae is complete,
the fraction of PNe which may at some time form a H-poor object, would
be between 5 and 10 per cent. However, if the H-poor phase has a much
shorter observability than a ordinary PN (as may be expected if less
material is ejected, and at higher velocity, than in the AGB ejection
phase), the final fraction may be higher. Bl\"ocker (2000) finds that
about 25\%\ of all central stars of PNe may experience a late thermal
pulse.  This is consistent with the assumption that all H-poor PNe
originate from such an event. However, it is not necessary to assume
that all late thermal pulses give rise to H-poor PNe.

If the timing of the late thermal pulse is random, a smaller fraction
of young PNe will show H-poor material. The observational difficulties
do not allow us to confirm this prediction based on the presently
available data.

The birth rate of PNe in the Galaxy is 0.5--1 yr$^{-1}$. Within our
Galaxy, a H-poor PN would form on the order of once per decade. 

\section{Members of the class}

\subsection{IRAS 15154$-$5258}

This PN was discovered by Manchado et al. (1989). The nebular diameter
is reported as 35 arcsec and the electron density $N_e = 10^2\,\rm
cm^{-3}$; the H-poor region has a diameter of 9 arcsec with
undetectable H$\alpha$ (Harrington 1996). The central star is
classified as [WC4]. There is considerable foreground extinction.
The images from which the diameters have been measured are unpublished.
The distance is unknown; given the properties, 5 kpc may be a reasonable
assumption.

\begin{figure}[H]
\centerline{\psfig{figure=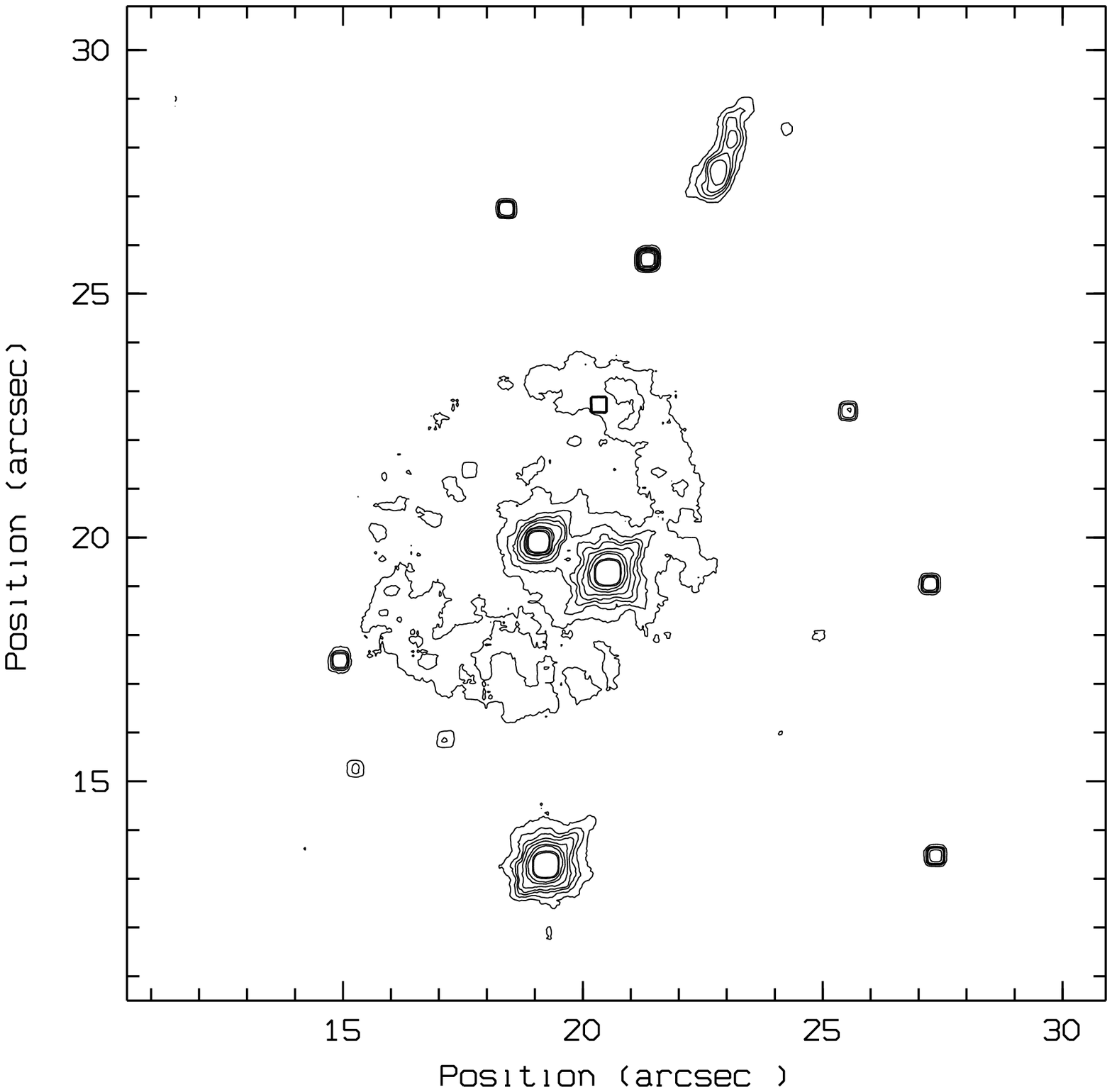,width=9cm}}
\centerline{\psfig{figure=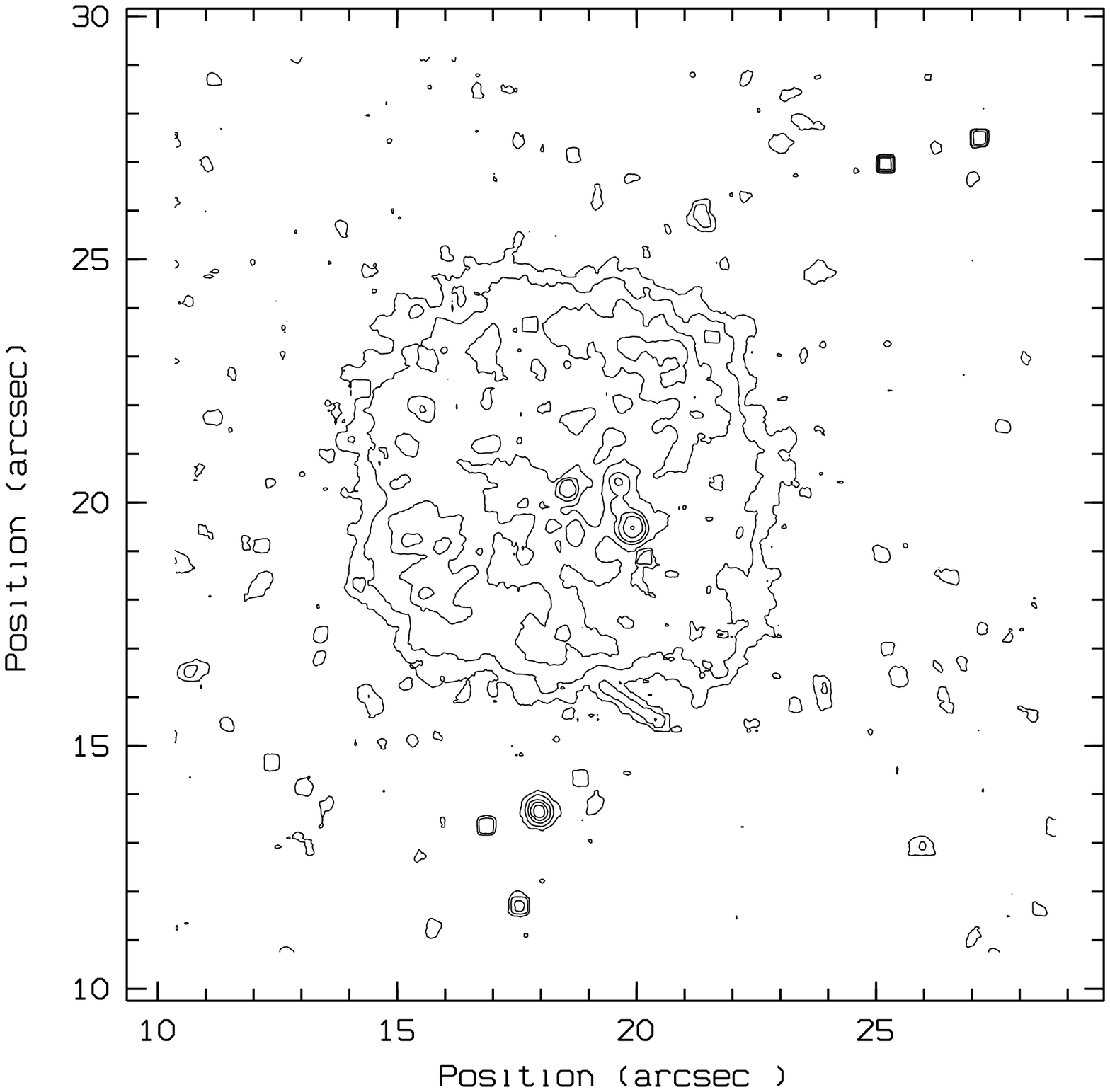,width=9cm}}
\caption{The HST H$\alpha$ and [OIII] images of IRAS 15154-5258}
\end{figure}

The HST images (taken with the Planetary Camera) are shown in Fig. 1.
The outer shell is not seen due to its faintness. The H$\alpha$ image
shows a limb-brightened shell surrounding the H-poor bubble. The
[OIII] image traces a slightly larger region but has much stronger
emission near the centre than seen in H$\alpha$.  Note the elongated,
jet-like feature visible in H$\alpha$. The [OIII] image shows strong
fagmentation of the shell, with elongated features pointing away from
the star, reminiscent of the cometary knots in the Helix nebula.

Manchado et al. comment on the strong IR emission and argue that the 
nebula is young. This is unlikely, given the very low density. The IR
emission is likely associated with the H-poor region instead.

\subsection{IRAS 18333$-$2357: the exception}

This PN is located in the Galactic globular cluster M22 (Gillet et
al. 1986, 1989). It is one of only 3 PNe known in globular clusters,
and one of 8--10 PNe known in the Galactic halo. To have such a rare
class represented in the halo population is remarkable but worrisome.
The distance of 3.2 kpc (Harris 1999, 1996) is accurately known, as is
the extinction ($E_{\rm B - V} = 0.34$) and the metallicity ([Fe/H]$ =
-1.64$), all derived from the parent cluster. The nebular diameter is
12 by 6 arcsec in [O III]; the density is $N_e = 10 \, \rm cm^{-3}$.
No hydrogen or helium lines have been detected in the nebula. The gas
mass is $m_{\rm g} \le 0.01\,\rm M_\odot$ and the dust mass $m_{\rm d}
\approx 8.4 \times 10^{-4} \,\rm M_\odot$.

The object is remarkable for several reasons. First, the morphology is
bow shaped (Borkowski et al. 1993), indicating the nebula is moving
through and interacting with the ISM. This is explained by the
velocity of the globular cluster: a similar effect is visible in K648,
the PN in the cluster M15 (Alves et al. 2000). Second,
photo-ionization heating is relatively ineffective at these low
densities: Borkowski \&\ Harrington (1991) show that instead the
electron gas is heated by photoelectric grain heating.  This is
an inefficient process which gives a very large ratio between
the IR (dust) luminosity and the emission-line luminosity.

Third, a spectrum of the central star (Harrington \&\ Paltoglou 1993)
shows substantial hydrogen and helium, as well as N-enrichment.  The
presence of hydrogen in the star is not easily reconciled with its
absence in the nebula. The luminosity of the star is also very high
for a PN central star with a low progenitor mass (the turn-off mass
of the cluster).

The peculiar location of the object, the presence of hydrogen in the
star and the high luminosity all suggests that this object is not a
true member of the class. Its origin may be sought in a binary merger
(as globular clusters are very dense star systems) rather than in a
late thermal pulse.

\subsection{A58: the youth}

A58 is a large, old PNe, with a bright edge on the southern side
indicating interaction with the ISM through which the star is
moving. The star underwent a nova outburst in 1919, and became a cool, R
Cor Bor-type star after which it ejected a dusty shell. After 1923,
the star became invisible due to circumstellar extinction from its
newly ejected dust shell. Pottasch et al. (1986) and Seitter (1987)
showed that these new ejecta were hydrogen-poor.  A58 therefore is the
living evidence that a late thermal pulse can lead to the ejection of
a hydrogen-poor nebula.  Pollacco et al. (1992) show velocities for
the ejecta of $\sim 200$ km/s: the red-shifted side is not seen due to
high extinction. NIR images (Zijlstra, unpublished) show that the
diameter of the dusty ejecta is 0.45 arcsec.  Mid-infrared ISO images
are presented by Kimeswenger et al. (1998). Interestingly, the star
shows a strong CIV line, seen in scattered light (Seitter 1987,
Pollacco, these proc.).  It should be classified as a [WC4] star because
of this.

Following its eruption, the star very quickly evolved first from the
WD cooling branch to the cool post-AGB branch, followed by reverse
evolution to a present temperature of $\sim 7 \, 10^4\,$K. The minimum
increase during the latter period was $10^3\,$K/yr.

From the diameter of the shell, one can estimate a mass $\sim 5 \times
10^{-4}\,\rm M_\odot$ if the density is $10^4 \, \rm cm^{-3}$. Seitter
(1987) estimates $10^{-2}\,\rm M_\odot$. The mass-loss episode, during
and following the 1923 disappearance act, must have reached at least
$10^{-5}\, \rm M_\odot\, yr^{-1}$ but may have been much higher. The
fact that the star is becoming visible again (in reflected light)
suggests that the mass loss rate has declined since.

\subsection{ A30 and A78: The elderly twins}

These are the easiest H-poor PNe to study: the ejections occured
probably $\sim 10^3\,$yr ago and the ejecta have expanded out to 10
arcsec from the stars. HST images (Borkowski et al. 1995, 1993) show
numerous cometary knots, embedded in a faster stellar wind.  The knots
are believed to have cold, neutral cores, and are dominated by
helium. The total amount of H-poor gas is $\sim 10^{-2}\,\rm M_\odot$.

Both A30 and A78 show evidence that the knots are located in a disk or
torus structure, with two additional knots located in the polar
direction, symmetric with respect to star.  The precise structures are
not identical (see the description in Harrington 1996) but the
similarities are striking. The polar knots are extraordinary well
aligned (to 5 arcmin), which may indicate that their origin is very
near the stellar surface. Dust emission in A30 is seen in the torus
but not in the polar knots (Dinerstein \&\ Lester 1984).

The outer nebulae have ages of order $10^4\,$yr. There is no question
that the stars evolved on normal post-AGB evolution before, at a late
stage, the H-poor material was ejected. This is consistent with a late
thermal pulse. However, Harrington (1996) argues that the disk/pole
structure is more akin to binary interaction and suggests the
evolution may have been more complicated. The outer nebulae are
spherical or mildly elliptical, often assumed to indicate  mass loss from
a single AGB star. 

The stars are hot, with temperatures of 1.1--1.2$\,10^5\,$K for both
A30 and A78. Borkowski et al. (1993) adopt a luminosity of 4000
L$_\odot$. If correct, this would indicate declining luminosities
compared to those expected shortly after a late thermal pulse.  Both
stars are classified as weak-emission-line stars (Tylenda et al.
1993). These may be related to the [WC] stars, but with weaker winds.
(The class is not well defined and probably contains H-rich objects
unrelated to the [WC] stars as well. The H-poor nature of the ejecta
makes a relation to the [WC] stars more likely.) The weaker wind could
be a consequence of a declining luminosity.

\subsection{Indicated evolution: A58$\rightarrow$IRAS1514$-$5258$\rightarrow$A30/A78}

Comparing the different objects shows that a late thermal pulse may
very quickly lead to hydrogen-poor eject in the centre of an old PN,
which may remain observable for at least $10^3\,\rm yr$. During this
time, the star makes a fast excursion to cool (7000 K) temperatures
but than heats up quickly. The sequence of A58 to A30/A78 (if this is
a correct interpretation) would indicate that after a rapid early
increase in temperature, subsequently the star evolves fairly slowly
back to the cooling branch with a slowly declining luminosity.

\section{Kinematics}

The outer nebulae show normal expansion velocities for old PNe, with
40 km/s for A78 (Meaburn et al. 1998) and A30 (Meaburn \&\ Lopez 1996)
and 31 km/s for A58 (Pollacco et al. 1992). The morphologies show only
small deviations from spherical symmetry, and the outer velocity
fields are probably also spherically symmetric, or mildly elliptical
(Meaburn \&\ Lopez 1996).

The inner edge of the outer nebulae shows clumpy, strongly enhanced
emission. For A30 and A78 this is located at a little over half of the
outer radius of the nebula (e.g. Harrington et al. 1995, par. 3),
while for IRAS 15154$-$5258 its radius is about 1/4 of the outer
radius.  These features are interpreted as a wind-swept shell, where
the present, fast and H-poor, wind from the star collides with the
H-rich gas in the older nebula. The stellar winds of A30 and A78 have
terminal velocities of 3600 km/s; the expansion velocity of the
swept-up shell is 73 km/s for A78 (Manchado et al. 1989, Harrington et
al. 1995), although highly variable with location (Meaburn et
al. 1998). The H-poor bubble inside this shell is therefore still
growing with respect to the outer nebula.

The knots described above are located inside this bubble. They are
moving much slower than the stellar wind, as shown by their morphology
(Borkowski et al. 1995). Meaburn et al. (1998) show that the knots in
the equatorial plane of A78 may be distributed throughout an elongated
disk, expanding at 25 km/s. The polar knots have much higher
velocities, of 380 km/s.  The kinematic age of the equatorial knots
may be similar to that of the swept-up shell, but the polar kots appear
to have much younger kinematic ages.

In A30 the central knots may show a similar disk/pole structure, but
due to the viewing angle these components are not as easy to
separate. Within the central regions, velocities of up to 200 km/s are
seen, which Meaburn \&\ Lopez (1996) attribute to material evaporating
of slow knots and mixing with the fast stellar wind. Such a
mass-loaded wind is also proposed for A78 (Harrington et al. 1995).

A58 has a very small hydrogen-poor region which is only barely
resolvable from the ground. Pollacco et al. (1992) show emission over
a velocity width of 360 km/s (assuming symmetry, since the red-shifted
emission is not seen due to dust absorption).  They suggest that the
flow may be collimated.

\section{Dust and abundances}

The hydrogen-poor regions are found to have a high fraction of dust.
In A58, the central star is almost completely obscured by dust in the
innermost region, which must have formed soon after the 1919 outburst
and caused the optical disappearance in 1923. IRAS 15154$-$5258 and
IRAS 18333$-$2357 also have strong IRAS emission. The dust colours
indicate that the dust is not very hot. Kimeswenger et al. (1998) find
$T_{\rm d} = 90\,$K for the dust in A58. The same authors show that in
the older nebula A78 the dusty region is located near the star,
possibly coincident with the knots.

The H-poor nebulae show a very large ratio of IR flux over optical
line flux. This suggests (Harrington 1996) that the heating of the
nebula is dominated by electrons ejected from the dust following
photon absorption.  Without this unusual heating source, the gas would
be too cool to be easily detectable.

Dust-to-gas ratios of the order of 0.2 (Borkowski \&\ Harrington 1991)
require both a large carbon overabundance and very efficient dust
formation.  Dust formation in carbon-rich regions is thought to be
initiated by $\rm C_2 H_2$. In H-poor regions, instead carbon chains
need to grow to provide the building blocks. LeToeff (2000) finds that
around hot stars, dust formation in such regions does not take place
unless extreme, unlikely densities are reached. Instead a cooler star
($\sim 10^4\, \rm K$ or less) is required. The condensation cores may
be formed by fullerenes ($\rm C_30$) which could form in a C-rich,
H-poor environment.

\begin{table}
\caption[]{Abundances by mass, as measured in the hydrogen-poor
regions}
\begin{tabular}{llllll}
\noalign{\smallskip}\hline\noalign{\smallskip}
     &  O/He  &  N/He  &  Ne/He &  C/He & \\
\noalign{\smallskip}\hline\noalign{\smallskip}
A30 polar knot & 0.0034   &  0.0020  &  0.0015  & 0.960: & Jacoby \&\ Ford 1983 \\
~~~equatorial knot     & 0.024    &  0.0070  &  0.014  & & Jacoby \&\ Ford 1983 \\
~~~equatorial plane     & 0.021    &  0.018   &  0.029   & & Kingsburgh \&\ Barlow 1994 \\
A78  & 0.021    &  0.0036  &  0.012  &     & Jacoby \&\ Ford 1983 \\
     & 0.740:    & 0.640:   &  0.210   &     & Manchado et al. 1988 \\      
18333$-$2357
    &  0.240     &        &  0.275  &     & Borkowski \&\ Harrington 1991 \\
\noalign{\smallskip}\hline\noalign{\smallskip}
\end{tabular}
\end{table}

Table 1 lists published nebular abundances. There is no detetectable
hydrogen and the abundances are therefore quoted with respect to
helium.  The numbers show substantial enrichment of all elements with
respect to helium. The polar knot shows lower abundances. Elements in
the dust (which will not include helium) are not included in the
listed values.

\section{Evolution}

\subsection{Relation to the [WC] stars}

It is natural to postulate a relation between the five H-poor PNe and
the H-poor [WC] central stars. The fact that two of the five have [WC]
central stars and two have weak-emission line stars indeed suggests a
direct link between the present objects and the [WC] stars.

The precise origin of the [WC] stars is under discussion.  The fact
that the late-thermal-pulse object A58 has a [WC] star shows that at
least some (probably most) of the [WC] stars originate from such an
event. However, this is probably not true for all [WC] stars. A group
of cool [WC] stars with extreme IR emission appears to have evolved
directly from the AGB (Zijlstra 2001). A search for H-poor material
around [WC] stars would be of interest (e.g. Pena et al. 2000).

If re-ignition of the helium occurs on the cooling track, 
the star will first return to the horizontal part of the Sch\"onberner
track. Later the luminosity will again decline slowly while the
helium burning tapers off (in contrast to the hydrogen-burning
stars where the luminosity drops very fast in the knee of the
Sch\"onberner track). It is tempting the associate the [WC]
star characteristics with the 'horizontal' part of the evolution and
the subsequent luminosity decline with the weak-emission-line stars.

\begin{figure}[t]
\centerline{\psfig{figure=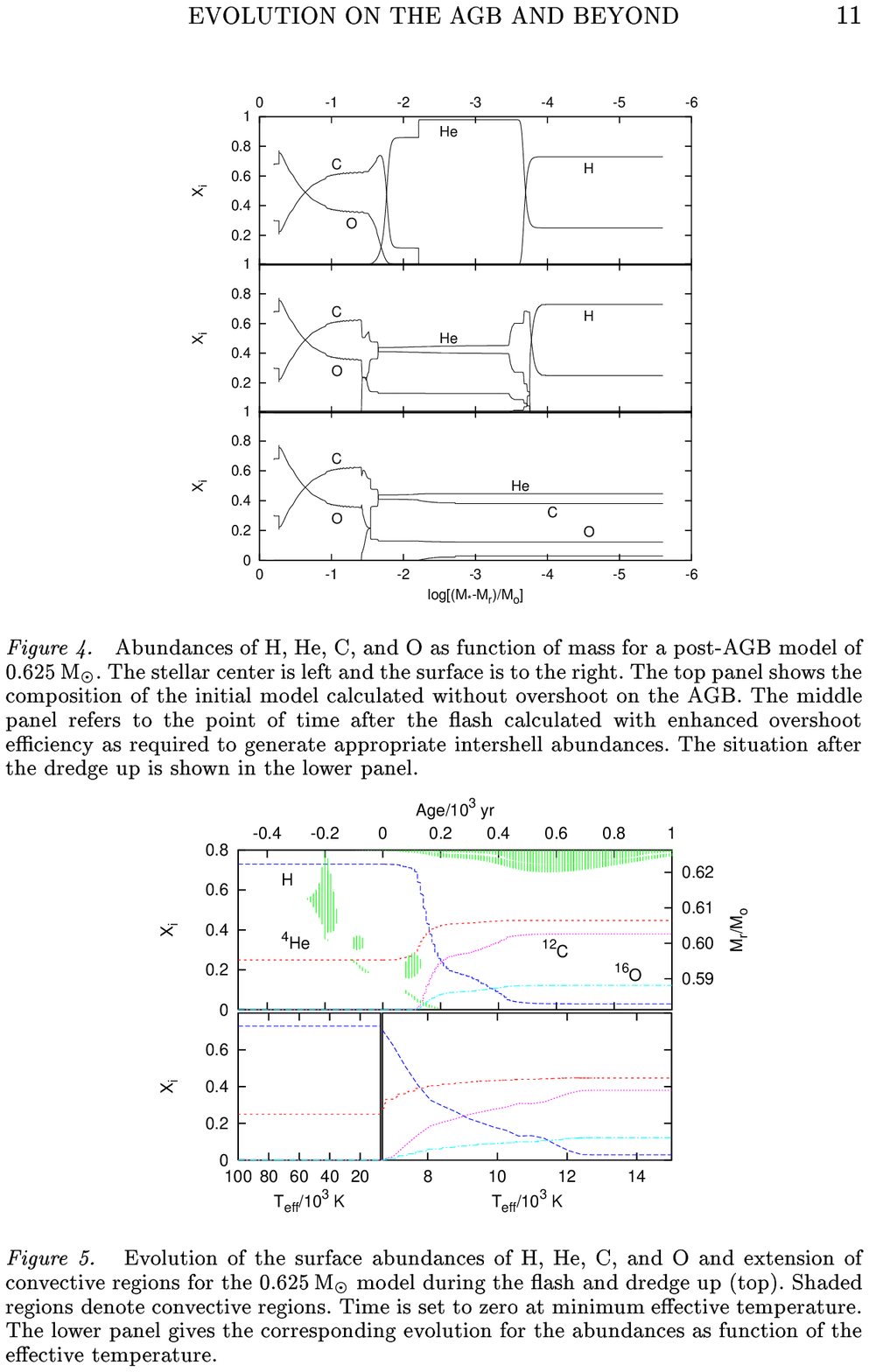,width=12cm,clip=}}
\caption{Atmospheric and sub-atmospheric abundances of a
post-AGB star (Bl\"ocker 2001). The panels show the mass fraction
versus the mass coordinate below the surface. The top
panel shows the abundances assuming no overshoot on the AGB.
The middle panel shows the abundances immediately after
the late thermal pulse, under the assumption of enhanced overshoot.
The bottom panel shows the same model after the dredge up which
follows the pulse.}
\end{figure}

\subsection{The helium layer}

The nebular abundances in Table 1 must arise from the upper layers of
the star. These layers are both nuclear processed and expelled
following the thermal pulse. The surface abundances of [WC] stars
should also approximate the abundances in the ejected, hydrogen-poor
material.

Typical surface abundances of [WC] stars (Dreizler \&\ Heber 1998) are
[He:C:O] = [0.33:0.50:0.17] by mass. Fig. 2 (upper panel) shows that
such abundance ratios do not occur anywhere in the upper layers of a
post-AGB star. Herwig et al. (1999) therefore propose that there is
diffusive overshoot into the helium layer, following thermal
pulses. This yields the modified abundance profiles shown in the
middle panel. During a late thermal pulse, the remaining hydrogen will
now be ingested into the helium layer and burned (bottom panel).
Prior to the late thermal pulse, the hydrogen-rich layer of the star
comprises about $10^{-4}\,\rm M_\odot$. The He layer contains a few
times $10^{-2}\,\rm M_\odot$.

The resulting abundances in the helium layer correspond well to the
[WC] stars: the large observed oxygen abundance requires efficient
diffusion. (But oxygen dredge-up in PNe does not support efficient
diffusion, Pequignot et al. 2000.)  Comparing the abundances in Table
1 shows that the H-poor PN also have an enhanced oxygen
abundance. However, the helium abundance appears to be much higher
than in the [WC] stars. Note that the carbon abundance is not well
known but could be high, especially if a large fraction is contained
in dust. The amount of H-poor gas in A30 and A78,  $\sim 10^{-2}\, \rm
M_\odot$, agrees with the mass of the helium layer.

The abundance ratios in the inner knots of A30 are in better agreement
with the non-diffusive overshoot model of the top panel, with the
exception of the uncertain carbon abundance. Better observed
abundances are needed: it appears that the H-poor PNe do not well
support the duffuse overshoot scenario. Instead it appears that the
surface abundances seen in [WC] stars are only reached after a
significant fraction of the helium layer has been ejected.  However,
the disappearance of hydrogen in Sakurai following its nova outburst
confirms the ingestion of hydrogen shortly after the thermal pulse.

The neon abundance in the intershell following a late thermal pulse is
of the order of 3.5\%. The extreme neon abundance in 18333$-$2357
cannot easily be explained and may confirm its origin in a separate
event.

\subsection{Mass loss history}

H-poor nebulae indicate substantial mass loss. A30 and A78 have
ejected $\sim 10^{-2} \, \rm M_\odot$ over $10^4\,\rm yr$, or an
average mass loss of $10^{-6}\,\rm M_\odot yr{-1}$, substantially
larger than their present winds of $\sim 10^{-9}\,\rm M_\odot
yr^{-1}$. The dense nebula in A58 shows that within a decade after the
pulse, mass loss rates of $10^{-5}\,\rm M_\odot yr^{-1}$ or (much)
higher can be reached.  

The models predict a short-lived luminosity spike coinciding with the
lowest temperature reached ($10^4\,\rm K$). This could trigger high
mass loss, which is not taken into account in the models.  The mass
loss may cause the star to evolve to much lower temperatures than
shown by the models.  The equatorial systems of knots in A30 and A78
have expansion velocities of $\sim 25\,$km/s: this is consistent with
the escape velocity of an AGB star ($\sim 3000\,$K).

A58 underwent a R Cor Bor-like phase within years after its late
thermal pulse. At this time its temperature was around 7000 K, and
episodic dust formation commenced. Shortly after the star disappeared.
The evolution of A58 may not have been identical to A30/A78, but it is
possible that the star of A58 resembled an AGB star while hiding
behind its dust shell, and experienced an unobserved superwind phase.

The disk-like structure of the knot system in A30 and A78 would also
have formed during this r-AGB phase ('r' for resemble).  The reason
for their morphology is not clear: one can speculate about magnetic
fields or back-fall of the nova ejecta.  After an equatorial system
formed, it could have collimated subsequent ejection events to give
the polar knots.  It would be of interest to see whether similar
evolution is also occuring in A58. Pollacco et al.  (1992) find some
indication for collimated flows, but the evidence is not conclusive.

The present temperature of the star of A58 is $\sim 7 \times 10^4\,$K
(based on its [WC4] class).  The excursion to low temperatures lasted
for no more than a few decade, possibly less. The bulk of the helium
layer may have been lost during a very short phase.

\subsection{Sakurai's object}

Thanks to A58, the association between Sakurai and the hydrogen-poor PNe
is not in doubt. We have a good idea how it will evolve over the next
$10^3\,\rm yr$, thanks to its older relatives. 

The immediate prediction from the evidence above is that Sakurai may
undergo a phase of very high mass loss. It is important to monitor its
evolution in the infrared, so that its mass-loss evolution can be
followed. The present wind velocity, from the He~I line, is 600 km/s
indicative of a still hot star. If the path of A30 and A78 is
followed, it is possible that the ejection velocity will drastically
reduce while the star becomes cooler.  The morphology may change to
form an equatorial disk. It would be important to test these
predictions, but the high extinction makes such studies very
difficult.

As more of the helium layer is removed, the abundance ratios in the
ejected material may change over time. As the expansion velocities
will vary over time, this tracer of the helium layer will become
more confused. Early studies of wind abundances are therefore important.

The dust-formation process can itself be studied for the first time.
But while shedding light on the dust formation in the absence of
hydrogen, the dust formation obscures our view of the evolution of the
underlying star. The study of Sakurai will be a challenging
opportunity for some time to come.

\end{article}
\end{document}